\documentclass[letterpaper,prl,twocolumn,superscriptaddress,showpacs]{revtex4}

\usepackage{graphicx}
\usepackage{bm}
\usepackage{amsmath}
\usepackage{color}

\newcommand{\rf}[1]{(\ref{#1})}

\newcommand{\beq}{\begin{equation}}
\newcommand{\eeq}{\end{equation}}
\newcommand{\beqr}{\begin{eqnarray}}
\newcommand{\eeqr}{\end{eqnarray}}
\newcommand{\lb}[1]{\label{#1}}
\newcommand{\bc}{\begin{center}}
\newcommand{\ec}{\end{center}}
\newcommand{\ct}[1]{\cite{#1}}
\newcommand{\bi}[1]{\bibitem{#1}}
\renewcommand{\section}[1]{{\par\it #1.---}}

\begin{document}

\title{Collective Effects in Nanolasers Explained by Generalized Rate Equations}

\author{I. E. Protsenko}
\affiliation{Lebedev Physical Institute of RAS, Leninsky Prospekt 53, Moscow, 119991, Russia}

\author{E. C. Andr{\'e}}
\affiliation{Department of Photonics Engineering, Technical University of Denmark, DK-2800 Kgs. Lyngby, Denmark}

\author{A. V. Uskov}
\affiliation{Lebedev Physical Institute of RAS, Leninsky Prospekt 53, Moscow, 119991, Russia}
\affiliation {ITMO University, Kronverksky Pr. 49, St. Petersburg, 197101, Russia}

\author{J. M{\o}rk}
\affiliation{Department of Photonics Engineering, Technical University of Denmark, DK-2800 Kgs. Lyngby, Denmark}

\author{M. Wubs}
\affiliation{Department of Photonics Engineering, Technical University of Denmark, DK-2800 Kgs. Lyngby, Denmark}


\begin{abstract}
\noindent We study the stationary photon output and statistics of
small lasers. Our closed-form expressions clarify the contribution of collective effects due to the interaction between quantum emitters.
We generalize laser rate equations and explain photon trapping: a decrease of the photon number output below the lasing threshold,
derive an expression for the stationary cavity mode autocorrelation function $g_2$, which implies that collective effects may strongly influence the photon statistics. We identify conditions for coherent, thermal and superthermal radiation, the latter being a unique fingerprint for collective emission in lasers. These generic analytical results agree with recent experiments, complement
numerical results, and provide insight into and design rules for nanolasers.
\end{abstract}

\pacs{42.50.Nn,
 78.67.Pt,
42.50.Ct
}

\maketitle

\noindent
Collective  effects (CE) in lasers, caused by the combined interaction of many quantum emitters through the lasing mode, are receiving increasing attention~\ct{Mlynek, Leymann, Jahnke, Kreinberg:2017a, Bohnet, Meiser, Temnov, Auffeves, Mascarenhas}. The collectively enhanced  dipole moments, as in free-space Dicke superradiance (SR)~\ct{Gross_Haroch}, increase the emitter-field coupling, and is reported to lower the lasing threshold~\ct{ Leymann, Meiser}, to  narrow the linewidth~\ct{Bohnet}, and to change the output photon statistics~\ct{Jahnke, Temnov, Kreinberg:2017a}, all beyond  standard laser theory. Qualitatively, by analogy with Dicke SR, one can expect the largest CE
in nano- and micro-lasers with a large number of emitters in a small volume, as in Q-dot nanolasers~\ct{Leymann} or VCSELs~\ct{Jahnke}, but more  quantitative predictions are much needed.

It is often supposed that
CE in lasers are due to high-order correlations and they are taken into account in various
sophisticated numerical approaches~\ct{Leymann, Auffeves}. Indeed, CE in nanolasers are more complex and still less understood than SR in free space. Analytical methods that provide further insight were developed mostly for low-quality cavities~\ct{Temnov} favorable for SR: the low-Q cavity provides a common radiative relaxation of emitters~\ct{Auffeves} similar to SR in free space.

In this Letter we present an analytical approach applicable not only to low-Q cavity lasers, which  gives new insight into CE in nanolasers, especially into their ``photon trapping''\ct{Leymann, Jahnke} and measured unusual photon statistics~\ct{Jahnke, Kreinberg:2017a}. We show explicit dependencies of these effects on the laser parameters.
%
\section{Basic equations}
\begin{figure}[t]
\bc \includegraphics[width=8.5cm]{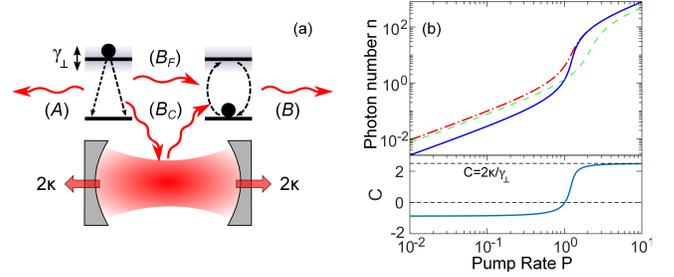}\ec
\caption{(Color online) Panel~(a): Sketch of the two-level laser model with emitter-emitter interaction in a cavity, compared to in free space. Besides via stimulated emission (not shown),
the de-excitation of an emitter is in two ways: ($A$) direct spontaneous emission, or ($B$) emission of a photon that excites another emitter, which subsequently emits a photon. The presence of both paths may increase the net emission rate, leading to SR, if the emitted photons are identical. For that process to occur with high probability in free space [marked by ($B_F$)], the emitters should be sub-wavelength spaced, while in a cavity, SR emission of photons may be due to the coupling of emitters to a common cavity mode [$(B_c)$], if the emitter linewidth $\gamma_{\perp} \leq 2\kappa$, the cavity linewidth.
Panel~(b): stationary photon output $n(P)$ for $ \kappa/\gamma_{\parallel} = 25$, $N_0 = 10^4$, $g = 0.048$, and $\gamma_{\perp}/\gamma_{\parallel} = 20$. Blue curve: GLRE solution based on Eq.~(\ref{70t}), with corresponding values of $C(P)$ at the bottom; green curve: same but without  CE ($C=0$); red curve: stationary solution of the LRE.
}\label{FIG1}
\end{figure}

%
We keep the analysis as simple as possible and consider a model of a single-mode laser with $N_0$ two-level emitters, shown in Fig.~\ref{FIG1}(a),
%
%
%
%
with transition frequencies equal to the lasing mode frequency~\ct{Leymann, Auffeves, Mascarenhas}. Following  \ct{Scully_book}, from the Hamiltonian~(S1) in the Supplemental Material SM1, we write the Maxwell-Bloch equations for the operators $\hat{a}$ of the cavity mode, the population inversion $\hat{\Delta}$, and the polarization $\hat{v} = \sum_{i=1}^{N_0}\hat{v}_i$, with $\hat{v}_i$  the  dipole operator of the $i^{\rm th}$ emitter:
\begin{subequations}\lb{2t}
\beqr
    \dot{\hat{a}} & = & \Omega_0\hat{v} - \kappa \hat{a} + \hat{F}_{a}  \lb{12}\\
    \dot{\hat{v}} & = & -(\gamma_{\perp}/2) \hat{v} + \Omega_0 f \hat{a}\hat{\Delta} + \hat{F}_{v}    \lb{13}\\
 \dot{\hat{\Delta}} & = & - 2\Omega_0(\hat{a}^+\hat{v} + \hat{v}^+\hat{a}) + \nonumber\\
   & + & \gamma_{\parallel}[P(N_0-\hat{\Delta}) - N_0-\hat{\Delta}] + \hat{F}_{\Delta}. \lb{14}
\eeqr
\end{subequations}
Here $\Omega_0$ is the vacuum Rabi frequency and $2\kappa$ is the decay rate of the cavity mode; the factor $f>0$ measures the average of squared couplings between emitters and the cavity mode;  $\gamma_{\perp}$ and $\gamma_{\parallel}$ are polarization and population relaxation rates;
$\gamma_{\parallel}P$ is the pump rate~\cite{Jesper}; the $\hat{F}_{\alpha}$ are the Langevin forces associated with the operators $\hat{\alpha} = \{\hat{a}, \hat{v}, \hat{\Delta}\}$. Below we denote expectation values of $\hat{\alpha}$ as $\alpha$.

Before discussing CE based on Eqs.~\rf{2t},  we recall that the usual  semi-classical Maxwell-Bloch equations~\ct{RE} are derived  by replacing all operators by c-number variables and by neglecting the  Langevin forces.  Then the stationary population inversion $\Delta = \Delta_{\rm th}={\gamma_{\perp}\kappa}/(2\Omega_0^2f)$ is the threshold one, and the cavity mode photon number $n \equiv |a|^2  = (2g)^{-1}(N_0/\Delta_{\rm th} + 1)({P}/P_{\rm th} - 1)$, where $g^{-1} = \gamma_{\parallel}\gamma_{\perp}/(4\Omega_0^2f)$ is the saturation photon number. Lasing requires a minimum number of  emitters $N_0>\Delta_{\rm th}$ and a minimal pump $P>P_{\rm th} = (N_0+\Delta_{\rm th})/(N_0-\Delta_{\rm th})$, giving a finite stationary cavity photon number $n>0$.

As the next standard approximation, the semiclassical laser rate equations for $n$ and $\Delta$~\ct{RE} can be
obtained from Eqs.~\rf{2t} by adiabatic elimination of the polarization $v$, which is valid at
$\gamma_{\perp} \gg \kappa, \gamma_{\parallel}$. Semiclassical Maxwell-Bloch or rate equations incorrectly predict $n=0$ at $P<P_{\rm th}$ since they do not take  spontaneous emission into the lasing mode into account. Laser rate equations (LRE) with spontaneous emission into the lasing mode are derived in~\ct{Rice}.
%
\section{Laser equations with collective effects}
%
Here we propose  generalized laser rate equations (GLRE) that are valid for {\em all} values of $\gamma_{\perp}$ since the polarization will not be integrated out, and include spontaneous emission into the lasing mode.  We take Eqs.~\rf{2t} as our starting point, neglect population fluctuations and replace  $\hat{\Delta}$ by its expectation value $\Delta$, a common approximation for weak coupling~\ct{Davidovich}. Thus Eqs.~\rf{2t} become linear in $\hat{a}$ and $\hat{v}$, and as a main result we derive the closed set of four equations (see SM2)
\begin{subequations}\lb{4t}\hspace{-1cm}\beqr
    \dot{{n}} & = & \Omega_0{\Sigma} - 2\kappa {n},   \lb{15}\\
    \dot{{\Sigma}} & = & -(\gamma_{\perp}/2 + \kappa){\Sigma} + 2\Omega_0 f( {n}{\Delta} + {N}_e + {D}),    \lb{16}\\
    \dot{{D}}  & = & -\gamma_{\perp} {D} + \Omega_0 {\Delta}{\Sigma}, \lb{17}\\
    \dot{{\Delta}} & = & - 2\Omega_0{\Sigma} + \gamma_{\parallel} [P(N_0-{\Delta}) - N_0-{\Delta}]. \lb{18}
\eeqr\end{subequations}
Here we introduce the collective emitter-field correlation ${\Sigma} \equiv \sum_{i=1}^{N_0}\left(\left<\hat{a}^+\hat{v_i}\right> + c.c\right)$ and  the collective dipole-dipole correlation ${D} \equiv f^{-1}\sum_{i\neq j}\left<\hat{v}_i^+\hat{v}_j\right>$, and use the operator identity $\hat{v}^+\hat{v} = f(\hat{N}_e + \hat{D})$,  where $\hat{N}_e$ is the operator for the population of upper lasing levels. Our GLRE~\rf{4t} reduces to the standard LRE [Eq.~(S7) of  SM2] at large dephasing~\ct{Siegman, Rice, Kurgin, Yamamoto} $2\kappa/\gamma_{\perp} \ll 1$. Below we show that collective effects are negligible in this limit.
%
\section{Stationary lasing with collective effects}
%
The stationary solution of Eqs.~\rf{4t} is obtained by putting the derivatives to zero, we do this in two stages: first we use the stationary Eqs.~\rf{16} and \rf{17} to express $\Sigma$ and $D$ in terms of $n$ and $\Delta$, this gives $D(n, \Delta) =  (n \Delta + N_e)C(\Delta)$, with
\beq
C(\Delta)  =  \frac{(2\kappa/\gamma_{\perp})(\Delta/\Delta_{\rm th})}{1 + (2\kappa/\gamma_{\perp})(1 - \Delta/\Delta_{\rm th})}. \lb{7aa}
\eeq
We identify $C(\Delta)$ as a dimensionless measure for the importance of CE (analogous to the cooperativity factor in \ct{Leymann}), which we now see to be small in the LRE limit. In the second stage, we insert the stationary $\Sigma(n, \Delta)$ and $D(n, \Delta)$ into the stationary Eqs.~\rf{15}, \rf{18}, and obtain
\begin{subequations}\lb{6t}\beqr
0 & = & G_{\Delta}(n \Delta + N_e) - (2\kappa/\gamma_{\parallel})n, \lb{8a}\\
0 & = & -2 G_{\Delta}(n \Delta + N_e) +
P(N_0 - \Delta) - N_0 - \Delta,  \lb{8b}
\eeqr\end{subequations}
where $G_{\Delta}\equiv g_{c} [1+C(\Delta)]$ with reduced coupling constant
$g_c \equiv g/(1 + 2\kappa/\gamma_{\perp})$. By comparison of the right-hand sides of Eqs.~(\ref{6t}) and of the LRE, it can be seen that stationary solutions of our GLRE model are obtained from the latter by the substitution $g \rightarrow G_{\Delta}$. By this variation of constants in a nonlinear model one can indeed expect qualitatively new behavior. The stationary photon output of Eqs.~(\ref{6t}) and hence of the GLRE of Eqs.~(\ref{4t})  is
\begin{subequations}\lb{70t}
\beqr
    n(P)  & = & \left[\theta(P) + \sqrt{\theta^2(P) +
    8{g_c}{P}N_0/\Delta_{\rm th}}\right]/(4 g), \lb{70t1}\\
    \theta(P) & = &  (P -1)N_0/\Delta_{\rm th} -  {P}  - 1 - g_c.   \lb{21a}
\eeqr\end{subequations}
At $2\kappa/\gamma_{\perp} \rightarrow 0$,
Eq.~\rf{70t} indeed converges to the known stationary solution of the LRE.

In Fig.~\ref{FIG1}(b) we show
 the photon output $n(P)$ based on Eq.~\rf{70t} for $2\kappa/\gamma_{\perp}=2.5$, for which the LRE limit is not valid. Nevertheless, for comparison we also show the corresponding LRE curve. The most distinct difference is a suppression in our generalized model of the photon output below threshold. (We return to this ``photon trapping'' shortly.) Above threshold, however, the LRE limit agrees surprisingly well with our GLRE predictions. This
follows from the fact that above threshold  $\Delta$ is clamped to $\Delta_{\rm th}$ and hence  $C(\Delta)$ to $2 \kappa/\gamma_{\perp}$ as shown by the horizontal dotted line on the bottom panel of Fig.~\ref{FIG1}(b), so that $G_{\Delta}$ tends to $g$ independent of the value of $\gamma_{\perp}$.
Thus our GLRE gives the simple but important prediction that for arbitrary values of $2\kappa/\gamma_{\perp}$,  the  stationary photon number above the lasing threshold is given by the standard LRE limit.
And yet CE play a role above threshold in the GLRE model, since  $C(\Delta)$ is clamped at a nonzero value, except in the LRE limit $2\kappa/\gamma_{\perp}\to 0$.

To further analyze the role of CE, in Fig.~\ref{FIG1}(b) we also present the photon output in the absence of CE, based on the stationary solutions~\rf{70t} with  $C(\Delta)=0$.
Compared to this curve, the most conspicuous collective effect of our full GLRE model are a suppression of the radiation at $\Delta < 0$
and an enhancement at $\Delta > 0$.
This photon trapping below threshold has been observed in numerical studies before~\ct{Leymann,Jahnke}, but never in a simple analytical model. We can correlate the `trapping'  to  $C(\Delta)$ being negative below threshold and the enhancement to $C(\Delta)> 0$  above, see the bottom of Fig.~\ref{FIG1}(b).

\begin{figure}[t]
\bc \includegraphics[width=8.5cm]{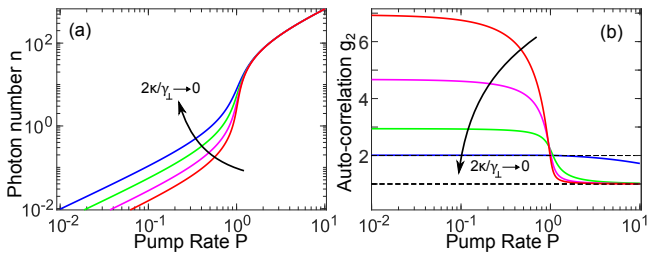} \ec
\caption{(Color online)
Panel~(a): Stationary photon output $n(P)$ for $2\kappa/\gamma_{\perp} = 0.01$ (blue), $1$ (green), $3$ (magenta), and $6$ (red); $g = 2/3$, and $\Delta_{\rm th}/N_0=0.01$.
Panel~(b): corresponding autocorrelation functions $g_2$ versus pump $P$.
}\label{FIG2}
\end{figure}

Fig.~\ref{FIG2}(a) is a systematic parameter study of
%
%
%
%
photon trapping in the model based on Eqs.~\rf{70t}. In particular, for lasers that
would exhibit the same semiclassical threshold $P_{\rm th}$, we compare the stationary photon output  as we vary the influence of CE by tuning $2\kappa/\gamma_{\perp}$.
Most discernible in Fig.~\ref{FIG2}(a) is again the photon trapping: for increasing $2\kappa/\gamma_{\perp}$,  the inter-emitter correlation $D$ increases and fewer photons are emitted below threshold. A corresponding    collective lasing enhancement above threshold is absent, since all curves tend to the same LRE limit, as explained above. The combined effect is that the lasing transition
becomes sharper as $2\kappa/\gamma_{\perp}$ is increased. In conventional theory, a sharper transition is related to a decreasing beta factor $\beta=g/(1+g)$~\ct{Rice}. Analogous to Ref.~\ct{Rice}, for our model we identify in SM3 the beta-factor $\beta_c \equiv g/(1+g + 2\kappa/\gamma_{\perp})$ that obviously decreases with $2\kappa/\gamma_{\perp}$, in correspondence with
the  threshold sharpening as CE become more  significant.
%
\section{Collective effects in laser statistics}
%
We now consider how in our model CE affect the stationary photon statistics of nanolasers, in particular the autocorrelation function $g_2 = \left<\hat{a}^+\hat{a}^+\hat{a}\hat{a}\right>/n^2$.
In usual laser theory, $g_{2}$ is sub-thermal, varying from 2 below threshold (thermal radiation) to  $1$ above (coherent). Super-thermal radiation, or "photon bunching" (with $g_2>2$)   has been  predicted to occur below threshold~\ct{Leymann, Auffeves} and measured both in pulsed~\ct{Jahnke} and in cw experiments~\ct{Kreinberg:2017a}. This $g_2>2$ constitutes a unique fingerprint of CE in lasers.

\begin{figure}[t]
\includegraphics[width=8.5cm]{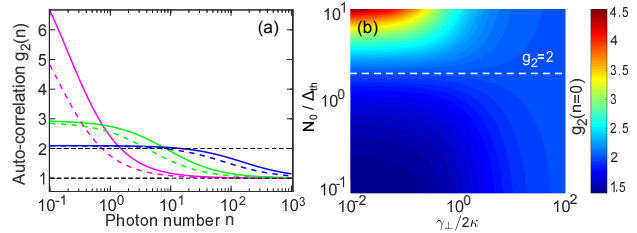}
\caption{(Color online) Photon autocorrelation function $g_2(n)$.
Panel (a): $g_2(n)$ for $N_0/\Delta_{\rm th} = 100$ (solid curves) and  $N_0/\Delta_{\rm th} = 50$ (dashed);   $\gamma_{\perp}/2\kappa = 0.1$ (magenta), 1 (green) and 10 (blue). Panel (b): $g_2(n=0)$ upon variation of parameters $\gamma_{\perp}/(2\kappa)$ and $N_{0}/\Delta_{\rm th}$.}\label{FIG3}
\end{figure}

We calculate $g_2$ following a similar approach as for Eqs.~\rf{4t}: using the dynamics of Eqs.~\rf{2t}, we derive in SM4 the equations of motion for the {\em four-operator} product $G_2 \equiv \left<\hat{a}^+\hat{a}^+\hat{a}\hat{a}\right>$ and determine its stationary mean value, again neglecting population fluctuations and Langevin forces.
We find as a main result
\beq
    g_2 = 1 + \frac{(1+\gamma_{\perp}/2\kappa)(d_{th}^{-1} +1)}
    {3+6n(1 + 2\kappa/\gamma_{\perp}) + (\gamma_{\perp}/2\kappa)(d_{th}^{-1} +1)}.\lb{27}
\eeq
This $g_2$ depends on the  two dimensionless parameters $2\kappa/\gamma_{\perp}$, the threshold population inversion per emitter $d_{th} = \Delta_{\rm th}/N_0$ and decreases monotonically with~$n$. In the large-$n$ limit of strong pumping, we find $g_2 = 1$, as in conventional laser theory.
By contrast,  we predict from the same Eq.~(\ref{27}) that for small $n$ super-thermal photon statistics will occur, for a low semiclassical threshold ($\Delta_{\rm th}/N_0 \ll 1$) in combination with a low decoherence rate $\gamma_{\perp}/2\kappa \leq 1$. This main result is illustrated in Fig.~\ref{FIG2}(b) as function of pump and in Fig.~\ref{FIG3} as a function of $n$.
%
%
%
%
Figs.~\ref{FIG2}(b) and~\ref{FIG3}(a)  also show that curves with larger $g_{2}(0)$ values  approach the coherent limit $g_2(n) = 1$ for smaller values of and $P$ and $n$, respectively. Thus, the emitter-emitter interaction leading to large $g_2>2$ helps to reach the coherent emission in accordance with results of~\ct{Leymann}.
For larger $\gamma_{\perp}/2\kappa$, the super-thermal radiation at small $n$ is suppressed, and the same happens when increasing the laser threshold $\Delta_{\rm th}/N_0$.
Finally, Fig.~\ref{FIG3}(a) illustrates that
for large decoherence ($\gamma_{\perp}/2\kappa \gg 1$), we indeed recover the LRE limit where $1\leq g_2 \leq 2$.

Eq.~\rf{27} also tells us what values of $g_2$ can be reached at $N_0 < \Delta_{\rm th}$ when lasing does not occur. Then $n$ saturates: $n\rightarrow n_s$ at $P \rightarrow \infty$, with $n_s$ given by Eq.~(S6) of SM2. For  $N_0 \ll \Delta_{\rm th}$, the  $g_2(n=0)$ varies from $4/3$ to $2$ when sweeping $\gamma_{\perp}/2\kappa$ from small to large values.
Thus in our model, devices that do not lase at strong pumping do not emit superthermal photons at weak pumping.

From Eq.~(\ref{27}) we identify
$g_2(n=0)< 2 + 2\kappa/\gamma_{\perp}$
as a useful upper bound for photon bunching. For further insight, in Fig.~\ref{FIG3}(b) we show $g_2(n=0)$  upon variation of $\gamma_{\perp}/(2\kappa)$ and $N_{0}/\Delta_{\rm th}$, again illustrating our prediction of strongly superthermal radiation at low decoherence $\gamma_{\perp}/2\kappa \ll 1$ in combination with a low lasing threshold $\Delta_{\rm th}/N_0 \ll 1$. For $\Delta_{\rm th}/N_0 > 1/2$ we find sub-thermal  radiation ($1< g_2 < 2$), whether lasing occurs or not.

%
\begin{figure}[t]
\includegraphics[width=8.5cm]{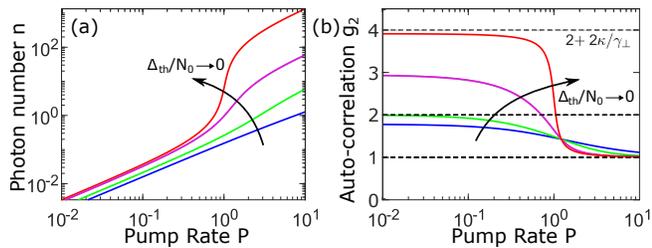}
\caption{(Color online) Pump dependence of photon output $n(P)$ [in panel (a)] and statistics $g_2(P)$ [in panel (b)], for fixed  $g = 2/3$ and $2\kappa/\gamma_{\perp} = 2$, while we vary  $\Delta_{\rm th}/N_0 = 0.9$ (blue), 0.5 (green), 0.1 (magenta), and 0.005 (red).}
\label{FIG4}
\end{figure}
%
%
Fig.~\ref{FIG2} illustrates that increasing $2\kappa/\gamma_{\perp}$ intensifies  emitter-emitter interaction leading to more photon trapping in~\ref{FIG2}(a) and, simultaneously, to more photon bunching in~\ref{FIG2}(b). Another way to obtain more strongly bunched photons, as shown in Fig.~\ref{FIG4}(b), is to decrease $\Delta_{\rm th}/N_0$ at some $2\kappa/\gamma_{\perp} > 1$. In doing so, the number of photons $n$ concomitantly  increases, see Fig.~\ref{FIG4}(a). However, due to collective photon trapping below threshold, the increase of $n$ is seen to be larger above than below the lasing threshold.

In SM5 we illustrate that our approach to identify generalized laser rate equations carries over to other lasing level schemes. We find that the level scheme affects the maximum value of $g_2$ that can be obtained. In particular, the scheme in SM5 where lower lasing levels depopulate infinitely fast  shows photon trapping but does not exhibit stationary superthermal photon statistics. Also,  the dependence of $g_2$ on the number of emitters is different.
%
\section{Discussion}
Our analytical main results~\rf{70t} for the photon output and~\rf{27} for their statistics show the conditions under which collective effects in lasers are significant.
The interaction between emitters through the cavity mode  increases with the reduction of the decoherence rate $\gamma_{\perp}$; a ``scale'' on which to compare this is the cavity linewidth $2\kappa$, and indeed we identified $2\kappa/\gamma_{\perp}\geq 1$ as a first  necessary condition for  strong  CE.

We obtain $\Delta_{\rm th}/N_0 \ll 1$ as the second necessary condition, or $12\pi fQN_0\gamma_{||}/(\gamma_{\perp}k^3V) \gg 1$, with $k$ and $Q$ the cavity mode wave number and quality factor. When neglecting pure dephasing, $\gamma_{\perp}=\gamma_{||}$ and the condition becomes $QN_0/(k^3V) \gg 1$. This is similar to the requirement $N_0/(k^3V) \gg 1$ for free-space SR for a large number of emitters in a volume $\sim k^{-3}$~\ct{Gross_Haroch}, and the difference is easily understood: in a cavity a photon interacts with emitters $Q$ times longer than in free space,  so SR in a cavity requires a $Q$ times smaller density.

Another important condition of SR in free space is that each emitter interacts with all others the same way, or ``feel the same environment''~\ct{Gross_Haroch}. This is approximately satisfied in a cavity, where all emitters interact  through the same cavity mode, so the variations in emitter-emitter interactions are of the order of fluctuations of atom-field couplings,
see SM1. Such fluctuations are small for large numbers of emitters.

We interpret the reduction of the mean photon number $n$ below threshold and the superthermal photon statistics as closely related manifestations of SR in the lasing cavity: photons are ``trapped'' by the emitter-emitter interaction, stored in the lasing medium for some time before being emitted in  groups.
During the storage phase, some photons are lost to nonradiative decay and to spontaneous emission to non-lasing modes, hence the reduced laser output. This picture agrees with measurements in other systems where SR pulses are delayed compared to single-emitter emission, with the more delayed pulses being less intense~\cite{Cong:2016a}.
Thus, provided the two necessary conditions for CE are   fulfilled,  below threshold a DC pumped lasing medium  emits a  random sequence of SR pulses into the lasing mode.

SR in free space is quite different for weak and for strong excitations: when  $N_0$ emitters
share one excitation, then the emission rate can be  $N_0$ times faster than for an  isolated emitter. On the other hand, when all $N_{0}$ emitters are excited, then they do not interact with each other and do not exhibit SR. Likewise, SR in the laser cavity can best be seen below threshold, at negative population inversion, as we showed.
%
\section{Conclusions and outlook}
We extended the standard laser rate equation model with two collective variables, describing  emitter-field and  emitter-emitter correlations. The resulting generalized laser rate equation  (GLRE) model can also describe existing lasers for which the polarization decays too slowly for the laser rate equations to be valid.
That is exactly the type of lasers where the usually neglected collective effects among the emitters can play a role. And indeed,
our simple model leads to analytical formulas~\rf{70t} and~\rf{27}, describing  observed photon trapping and superthermal photon bunching, both interpreted as consequences of superradiance in lasers.

Superthermal photon bunching is the most interesting effect,
characterized by the second-order autocorrelation function $g_2>2$, whereas $g_2=2$ is the upper limit for cw pumping in standard laser theory. We predict the maximally bunched light for  $2\kappa/\gamma_{\perp} \gg 1$ in combination with a low lasing threshold. The photon bunching occurs only below
threshold.

Our results are in qualitative agreement with those of more sophisticated numerical models~\ct{Leymann} and~\ct{Jahnke}. That photon trapping and superthermal bunching are both already captured in our model, where correlations beyond fourth order are neglected, is one of our surprising findings.  Our  approach can be
generalized beyond two-level systems, as we illustrated in SM5, and is likely to find further applications in the quantum theory of nanolasers, thereby substantially advances the fields of nanolaser physics and driven quantum systems.

More practically, our results give design rules for engineering the unusual stationary quantum statistical properties of radiation from a single laser cavity. For complementary schemes to engineer dynamical properties with coupled cavities, see~\ct{Marconi}. Cavity-enhanced superbunched light in a well-defined mode may find applications in the second- and higher-order interference of light and high-visibility ghost imaging with classical light~\ct{Zhou}.
%
\begin{acknowledgments}
We thank Mikkel Settnes for stimulating discussions.
This work benefited from the COST Action
MP1403
AU was supported by the ITMO Visiting Professorship program (Grant 074-U01). IP and AU was supported within the PRS CNRS/RFBR (Grant RFBR-17-58-150007). ECA, JM, and MW acknowledge support from the NATEC Centre funded by VILLUM FONDEN (grant 8692).
\end{acknowledgments}


\begin{thebibliography}{10}

\bi{Mlynek} J.~A.~Mlynek, A.~A.~Abdumalikov, C.~Eichler, and A.~Wallraff, {\em Observation of Dicke superradiance for two artificial atoms in a cavity with high decay rate}, Nature Commun. {\bf 5}, 5186 (2014).

\bi{Leymann} H.~A.~M.~Leymann, A.~Foerster, F.~Jahnke, J.~Wiersig, and C.~Gies, {\em Sub- and superradiance in nanolasers}, Phys. Rev. Appl. {\bf 4},  044018 (2015).

\bi{Jahnke} F.~Jahnke, G.~Gies, M.~A{\ss}mann, M.~Bayer, H.~A.~M.~Leymann, A.~Foerster, J.~Wiersig, C.~Schneider, M.~Kamp, and S.~H{\"o}fling, {\em Giant photon bunching, superradiant pulse emission and excitation trapping in quantum-dot nanolasers}, Nature Commun. {\bf 7}, 1540 (2016).

\bi{Kreinberg:2017a} S.~Kreinberg, W.~W.~Chow, J.~Wolters, C.~Schneider, C.~Gies, F.~Jahnke, S.~H{\" o}fling, M.~Kamp, and S.~Reitzenstein, {\em Emission from quantum-dot high-$\beta$ microcavities: transition from spontaneous emission to lasing and the effects of superradiant emitter coupling}, Light: Science $\&$ Appl.  {\bf 6}, e17030 (2017).

\bi{Bohnet} J.~G.~Bohnet, Z.~Chen, J.~M.~Weiner, D.~Meiser, M.~J.~Holland, and J.~K.~Thompson, {\em A steady-state superradiant laser with less than one intracavity photon},   Nature  {\bf 484}, 78 (2012).

\bi{Meiser} D.~Meiser and M.~J.~Holland, {\em Steady-state superradiance with alkaline-earth-metal atoms}, Phys. Rev. A {\bf 81}, 033847 (2010).

\bi{Temnov} V.~V.~Temnov and U.~Woggon, {\em Photon statistics in the cooperative spontaneous emission}, Opt.  Express {\bf 17} 5774 (2009).

\bi{Auffeves} A.~Auff{\`e}ves, D.~Gerace, S.~Portolan, A.~Drezet, and M.~F.~Santos, {\em Few emitters in a cavity: from cooperative emission to individualization}, New J. Phys. {\bf 13} 093020 (2011).

\bi{Mascarenhas} E.~Mascarenhas, D.~Gerace, M.~F.~Santos, and A.~Auff{\`e}ves, {\em Cooperativity of a few quantum emitters in a single-mode cavity}, Phys. Rev. A {\bf 88}, 063825 (2013).

\bi{Gross_Haroch} M.~Gross and S.~Haroche, {\em Superradiance: an essay on the theory of collective spontaneous emission}, Phys. Rep. {\bf 93}, 301 (1982).

\bi{Scully_book} M. O. Scully and M. S. Zubairy, {\em Quantum Optics} 	(Cambridge University Press, 1997).

\bi{Jesper} A.~Moelbjerg, P.~Kaer, M.~Lorke, B.~Tromborg, and J.~M{\o}rk, {\em Dynamical properties of nanolasers based on few discrete emitters}, IEEE J. Quant. Electron. {\bf 49} 945 (2013).

\bi{RE} M.~Sargent~III, M.~O.~Scully, and W.~E.~Lamb~Jr., {\em Laser Physics} (Addison-Wesley, New York, 1993).

\bi{Rice} P.~R.~Rice and H.~J.~Carmichael, {\em Photon statistics of a cavity-QED laser: A comment on the laser-phase-transition analogy}, Phys. Rev. A {\bf 50}, 4318 (1994).

\bi{Davidovich} I.~Protsenko, P.~Domokos, V.~Lef{\`e}vre-Seguin, J.~Hare, J.~M.~Raimond, and L.~Davidovich, {\em Quantum theory of a thresholdless laser},  Phys. Rev. A {\bf 59}, 1667 (1999).

\bi{Knight} P.~W.~Milonni and P.~L.~Knight, {\em Retardation in the resonant interaction of two identical atoms}, Phys. Rev. A {\bf 10}, 1096 (1974).

\bi{Siegman}  A.~E.~Siegman, {\em Lasers} (University Science Books; Sausalito, CA, 1986).

\bi{Kurgin} J.~B.~Khurgin and G.~Sun, {\em Comparative analysis of spasers, vertical-cavity
surface-emitting lasers and surface-plasmon-emitting diodes}, Nature Photon. {\bf 8},  468 (2014).

\bi{Yamamoto} G.~Bj{\"o}rk, A.~Karlsson, and Y.~Yamamoto, {\em Definition of a laser threshold}, Phys. Rev. A {\bf 50}, 1675 (1994).

\bibitem{Cong:2016a} K. Cong, Q. Zhang, Y. Wang, G.T. Noe II, A. Belyanin, and J. Kono, {\em Dicke superradiance in solids [Invited]}, J.~Opt. Soc. Am. B {\bf 33}, C80 (2016).

\bi{Marconi} M.~Marconi, J.~Javaloyes, P.~Hamel, F.~Raineri, G.~Beaudoin, I.~Sagnes, A.~Levenson and A.~M.~Yacomotti, {\em Quench dynamics in strongly coupled laser cavities}, ArXiv:1706.02993v3 [physics.optics] (2017).

\bi{Zhou}Y.~Zhou, Fu-li~Li, B.~Bai, H.~Chen, J.~Liu, Z.~Xu, and H.~Zheng, {\em Superbunching pseudothermal light}, Phys. Rev. A {\bf 95}, 053809 (2017).

\end{thebibliography}
\end{document}


\title{Collective Effects in Nanolasers Explained by Generalized Rate Equations}

\author{I. E. Protsenko}

\author{E. C. Andr{\'e}}

\author{A. V. Uskov}

\author{J. M{\o}rk}

\author{M. Wubs}


\maketitle

\section{{\bf Supplementary material (SM):}}
\subsection{SM1: Maxwell-Bloch operator equations}
\noindent We consider a single-mode laser with $N_0$ two-level emitters interacting resonantly with the lasing mode. Denoting the vacuum Rabi frequency of the lasing mode by $\Omega_0$ and using the rotating wave approximation (RWA), the Hamiltonian of the system can be written as
\beq
    \hat{H} = -\hbar\Omega_0\sum_{i=1}^{N_0}f_i(\hat{a}^+\hat{\sigma}_i + \hat{\sigma}_i^+\hat{a}) + \hat{\Gamma}, \lb{H1}
\eeq
where $\hat{\sigma}_i$  is the lowering  operator for $i$'th emitter, $\hat{a}$ is the cavity mode annihilation operator, $f_i$ quantifies the coupling between the $i$'th emitter and the cavity mode, and $\hat{\Gamma}$ describes the interactions with external reservoirs. For example, for a Fabry-P{\'e}rot cavity $f_i = \sin{(kr_i)}$, where $k = \omega_0/c$ is the wave number of the cavity mode.

From the Hamiltonian \rf{H1} we can derive equations describing the dynamics of the operators as detailed in~[11]:
\begin{subequations}\lb{1t}
\beqr
    \dot{\hat{a}} & = & i\Omega_0\sum_{i=1}^{N_0}f_i\hat{\sigma}_i - \kappa \hat{a} + \hat{F}_{a}, \lb{1}\\
    \dot{\hat{\sigma}}_i & = & -(\gamma_{\perp}/2)\hat{\sigma}_i - i\Omega_0f_i(\hat{n}_i^e-\hat{n}_i^g)\hat{a} + \hat{F}_{\sigma_i}, \lb{2} \\
    \dot{\hat{n}}_i^e & = & \gamma_{\parallel} (P\hat{n}_i^g - \hat{n}_i^e) - \nonumber\\
& & \hspace{1.5cm}
    i\Omega_0f_i(\hat{a}^+\hat{\sigma}_i - \hat{\sigma}_i^+\hat{a}) + \hat{F}_{n_i^e}, \lb{3} \\
    \dot{\hat{n}}_i^g & = & \gamma_{\parallel} (\hat{n}_i^e - P\hat{n}_i^g) + \nonumber\\
& &  \hspace{1.5cm}  i\Omega_0f_i(\hat{a}^+\hat{\sigma}_i - \hat{\sigma}_i^+\hat{a}) + \hat{F}_{n_i^g}. \lb{4}
\eeqr\end{subequations}
Here $\hat{n}_i^e = \hat{\sigma}_i^+\hat{\sigma}_i$ ($\hat{n}_i^g = 1 - \hat{n}_i^e$) is the population operator corresponding to the excited (ground) state of the $i$'th emitter, and $\hat{F}_{\alpha}$ is the Langevin force operator associated with the operator $\hat{\alpha}=\left\{\hat{a},\hat{\sigma}_i,\hat{n}_i^e,\hat{n}_i^g\right\}$. The population relaxation rate $\gamma_{\parallel}$ quantifies the non-radiative decay rate and the rate of spontaneous emission into non-lasing modes, $2\kappa$ is the lasing mode photon decay rate, and the rate $\gamma_{\perp}$ describes the decay of the polarization. The polarization decay rate depends on the pump rate $\gamma_{\parallel}P$ via
\beq
	\gamma_{\perp} = 2\gamma_{d} + \gamma_{\parallel}(1+P) \lb{0140}
\eeq
as in Ref.~[12], where $\gamma_{d}$ is the pure dephasing rate. When calculating the stationary mean intra-cavity photon number $n$,  we assume for simplicity that $2\gamma_{d} \gg \gamma_{\parallel}P$ and neglect the dependence of $\gamma_{\perp}$ on $P$. The relaxation terms and Langevin forces are added to equations \rf{1t} by the standard procedure describing the interaction with baths in the Markov approximation~[11].

Finally, we introduce operators for macroscopic quantities: The total polarization $\hat{v} = i\sum_{j=1}^{N_0}f_j\hat{\sigma}_j$, the total populations of the excited and ground states $\hat{N}_{e,g} = \sum_{i=1}^{N_0}\hat{n}_i^{e,g}$, the total population inversion $\hat{\Delta} = \hat{N}_e-\hat{N}_g$ and the Langevin forces $\hat{F}_{v} = \sum_{j=0}^{N_0}\hat{F}_{v_j}$, $\hat{F}_{v_j} = if_j\hat{F}_{\sigma_j}$, $\hat{F}_{\Delta} = \sum_{i=0}^{N_0}(\hat{F}_{n_i^e} - \hat{F}_{n_i^g})$. Then, approximating $f_i^2 \approx f \equiv N_0^{-1}\sum_{i=1}^{N_0}f_i^2$ and summing Eqs.~\rf{1t} over all emitters we obtain Eqs.~(1) of the main text.
%
%
\subsection{SM2: Laser equations with collective effects}
%
%
\noindent Here we will derive Eqs.~(2) of the main text that define our GLRE model. Using Eqs.~\rf{1t} and~(1) we can derive four equations for four coupled variables: The photon number operator $\hat{n} = \hat{a}^+\hat{a}$, the emitter-field interaction $\hat{\Sigma} = \hat{a}^+\hat{v} + \hat{v}^+\hat{a}$, the total population inversion $\hat{\Delta}$, and finally the  product $\hat{D}_t = f^{-1}\hat{v}^+\hat{v}$ of polarization creation and annihilation operators. Terms $\left<\hat{a}^+\hat{v_i}\right>$ and $\left<\hat{v}_i^+\hat{v}_j\right>$ in $\Sigma$ and $D$ are analogous, respectively, to the photon-assisted transition amplitude $P_X^{\alpha}$ and to the superradiant coupling correlation function $C_{\alpha\beta}^x$ introduced in~[2] with $\alpha, \beta$ standing for $i$, $j$. The four equations are
\begin{subequations}\lb{04t}\hspace{-1cm}\beqr
    \dot{\hat{n}} & = & \Omega_0\hat{\Sigma} - 2\kappa \hat{n} + \hat{F}_n,  \lb{015}\\
    \dot{\hat{\Sigma}} & = & -(\gamma_{\perp}/2 + \kappa)\hat{\Sigma} + \nonumber\\ & & \hspace{1.5cm} 2\Omega_0 f(\hat{\Delta} \hat{n}+ \hat{D}_t) + \hat{F}_{\Sigma},   \lb{016}\\
    \dot{\hat{D}}_t  & = & -\gamma_{\perp} \hat{D}_t + \Omega_0(\hat{a}^+\hat{\Delta}\hat{v} + \hat{v}^+\hat{\Delta}\hat{a}) + \hat{F}_{D_t}, \lb{017}\\
    \dot{\hat{\Delta}} & = & - 2\Omega_0\hat{\Sigma} + \nonumber\\ & & \hspace{0.8cm} \gamma_{\parallel} [P(N_0-\hat{\Delta}) - N_0-\hat{\Delta}] + \hat{F}_{\Delta}, \lb{018}
\eeqr\end{subequations}
where
\begin{align*}
    \hat{F}_n &= \hat{a}^+\hat{F}_a + \hat{F}_{a^+}\hat{a}, \\
		\hat{F}_{\Sigma} &= \hat{a}^+\hat{F}_v  + \hat{v}^+\hat{F}_a + h.c.\\
		\hat{F}_{D_t} &= f^{-1}(\hat{F}_{v^+}\hat{v} + \hat{v}^+\hat{F}_v).
\end{align*}
 Note that we preserve the order of non-commuting operators $\hat{\Delta}, \hat{v}$ and $\hat{v}^+$ in Eq.~\rf{017}.
Using the operator relation $(\hat{n}_j^e - \hat{n}_j^g)\hat{\sigma}_i = -\hat{\sigma}_i\delta_{ij}$ it is seen that
\[
    \hat{a}^+\hat{\Delta}\hat{v} + \hat{v}^+\hat{\Delta}\hat{a} = \hat{a}^+(\hat{\Delta}\hat{v})' + (\hat{v}^+\hat{\Delta})'\hat{a}
    - (\hat{a}^+\hat{v} + \hat{v}^+\hat{a}),
\]
where
\[ (\hat{\Delta}\hat{v})' = \sum_{k\neq j}(\hat{n}_k^e - \hat{n}_k^g)if_j\hat{\sigma}_j, \]
 with the prime denoting that operator products of the same emitters are excluded from sums in $\hat{v}^+\hat{v}$, so that
$(\hat{\Delta}\hat{v})' = (\hat{v}\hat{\Delta})'$. If we assume $N_0 \gg 1$, and neglect fluctuations of $\hat{\Delta}$, replacing $\hat{\Delta}$ by its expectation value $\Delta$, we can approximate $(\hat{\Delta}\hat{v})' \approx \Delta\hat{v}$ and $(\hat{v}^+\hat{\Delta})' \approx \Delta\hat{v}^+$, with the precision $1/N_0 \ll 1$. In this case,
\[
    \hat{a}^+\hat{\Delta}\hat{v} + \hat{v}^+\hat{\Delta}\hat{a} \approx (\Delta - 1)\hat{\Sigma}.
\]
Inserting this and $\hat{\Delta} = \Delta$ into Eqs.~\rf{04t}, we find the following equations for the expectation values of the operators:
\begin{subequations}\lb{041t}\hspace{-1cm}\beqr
    \dot{n} & = & \Omega_0\Sigma - 2\kappa n,   \lb{0151}\\
    \dot{\Sigma} & = & -(\gamma_{\perp}/2 + \kappa)\Sigma + 2\Omega_0 f(n \Delta + {D}_t),  \lb{0161}\\
    \dot{D}_t  & = & -\gamma_{\perp} D_t + \Omega_0 (\Delta-1)\Sigma \nonumber\\ & & \hspace{2.2cm} + \gamma_{\parallel}PN_g + (\gamma_{\perp}- \gamma_{\parallel})N_e, \lb{0171}\\
    \dot{\Delta} & = & - 2\Omega_0\Sigma +  \gamma_{\parallel} [P(N_0-\Delta) - N_0-\Delta]. \lb{0181}
\eeqr\end{subequations}
Here it has been used that $\left<\hat{F}_n\right> = \left<\hat{F}_{\Sigma}\right> = \left<\hat{F}_{\Delta}\right> = 0$;
moreover, it has been used that the diffusion coefficient~[11]
\[
	2D_{v^+ v}= \left<\hat{v}^+\hat{F}_{v}\right> + \left<\hat{F}_{v^+}\hat{v}\right>,
\]
gives
\[
	\left<\hat{F}_{D_t}\right> = 2D_{v^+v} / f= \gamma_{\parallel}PN_g + (\gamma_{\perp}- \gamma_{\parallel})N_e.
\]
Now, using the operator relation $\hat{\sigma}_j^+\hat{\sigma}_j = n_j^e$, we replace $D_t = D + N_e$, where
\[
	{D} = f^{-1}\sum_{j\neq k}f_j f_k\left<\hat{\sigma}_j^+\hat{\sigma}_k\right>
\]
describes the inter-emitter correlation. By summing Eq.~\rf{3} over all emitters and taking  the expectation value we find
\[
	\dot{N}_e = \gamma_{\parallel} (PN_g - N_e) - \Omega_0\Sigma,
\]
so
\begin{align*}
    \dot{D}_t  &=  -\gamma_{\perp} D + \Omega_0 (\Delta-1)\Sigma + \gamma_{\parallel}PN_g - \gamma_{\parallel}N_e\\
		&=-\gamma_{\perp} D + \Omega_0 \Delta\Sigma + \dot{N}_e,
\end{align*}
which, together with $\dot{D}_t - \dot{N}_e = \dot{D}$, implies that Eqs.~\rf{041t} indeed lead to the four coupled GLRE equations~(2). Consequently, the stationary photon number $n$ is given by Eqs.~(5) of the main text. When the number $N_0$ of emitters that is so small that the threshold population inversion per emitter $\Delta_{\rm th}/N_0 > 1$ and lasing is impossible, then one can see from Eq.~(5b) that $\theta(P) < 0$ for all $P$, and that the cavity photon number $n(P)$ saturates at $P\rightarrow \infty$:
\beq
n(P)\rightarrow n_s = [(\Delta_{\rm th}/N_0 - 1)(1+2\kappa/\gamma_{\perp})]^{-1}. \lb{800}
\eeq
The GLRE~(2) reduce to the standard laser rate equations in case of large pure dephasing. In more detail,
when $\gamma_{\perp} \gg \kappa, \gamma_{\parallel}$
we can adiabatically eliminate $\Sigma$ and $D$ from Eqs.~(2b) and~(2c)
by setting  $\dot{\Sigma} =\dot{D} = 0$,
and thereby express $\Sigma$ and $D$ in terms of $n$ and $\Delta$. Using this in Eqs.~(2a) and~(2d) and neglecting  terms $\sim \kappa/\gamma_{\perp} \ll 1$ we obtain the laser rate equations (LRE)
\begin{subequations}\lb{3t}\beqr
    \gamma_{\parallel}^{-1}\dot{n} & = & g(n \Delta  + N_e)- (2\kappa/\gamma_{\parallel})n,   \lb{3t1}\\
    \gamma_{\parallel}^{-1}\dot{\Delta} & = & - 2g(n \Delta  + N_e)+  \nonumber\\
    & & \hspace{1.5cm} P(N_0-\Delta) - N_0-\Delta. \lb{3t2}
\eeqr\end{subequations}
Here $g N_e$ is the dimensionless rate of spontaneous emission into the lasing mode in units of $\gamma_{\parallel}$.

The laser rate equations of Eqs.~\rf{3t} only valid in the limit $2\kappa/\gamma_{\perp} \ll 1$, when collective effects are suppressed. Yet for  low-Q cavity micro- and nano-lasers such as VCSELs, values of $2\kappa/\gamma_{\perp} \geq 1$ are not uncommon~[2].
%
\subsection{SM3: Factors $\beta$, $\beta_c$}
%
\noindent Here our aim is to identify a beta-factor $\beta_{c}$ for our GLRE model with collective effects, in analogy with the beta factor $\beta$ of the common LRE model. To start with the latter, we rewrite Eq.~\rf{3t2} as
\[
    \gamma^{-1}_{\parallel}\dot{\Delta} = - 2g n \Delta +   P(N_0-\Delta) - (1+g)(N_0+\Delta),
\]
using $N_e=\tfrac{1}{2}(N_0 + \Delta)$. We then re-normalize $\gamma_{\parallel}$ replacing it by $\gamma_{\rm tot} = \gamma_{\parallel}(1+g)$, which is the population relaxation rate taking into account spontaneous emission into the lasing mode. Then we rewrite Eqs.~\rf{3t}
\begin{subequations}\lb{31t}\beqr
    \gamma_{\rm tot}^{-1}\dot{n} & = & \beta(n \Delta + N_e) - (2\kappa/\gamma_{\rm tot})n,   \lb{31t1}\\
    \gamma_{\rm tot}^{-1}\dot{\Delta} & = & - 2\beta n \Delta + P'(N_0-\Delta) - N_0-\Delta, \lb{31t2}
\eeqr\end{subequations}
where $P' = P\gamma_{\parallel}/\gamma_{\rm tot}$ and $\beta = g/(1+g)$ is the fraction of spontaneous emission going into the lasing mode.

Analogously, we may introduce the parameter $\beta_c$ for the generalized laser rate equations~(2). We introduce the population relaxation rate $\gamma_c = \gamma_\parallel(1+g_c)$, which includes the spontaneous emission rate into the lasing mode from non-interacting emitters (but does not include the rate of collective spontaneous emission), and we re-write Eqs.~(4) of the main text
\begin{subequations}\lb{60t}\beqr
0 & = & \beta_c(n \Delta + N_e + D) - (2\kappa/{\gamma_{\parallel}})n, \lb{80a}\\
0 & = & -2\beta_c(n \Delta  + D) +\nonumber\\& & \hspace{1cm} P_c(N_0 - \Delta) - N_0 - \Delta,  \lb{80b}
\eeqr\end{subequations}
where $\beta_c = g_c/(1+g_c) = g/(1+ 2\kappa/\gamma_h + g)$ and $P_c = P\gamma_{\parallel}/\gamma_c$;
it is clear that $\beta_c$ decreases monotonically as a function of $(2\kappa/\gamma_{\perp})$. The important finding is that collective effects reduce spontaneous emission into the lasing mode, and this agrees with our finding in the main text that they make the lasing transition sharper, see  Fig.~2(a) of the main text.
%
\subsection{SM4: Derivation of equation~(6)}
%
From Eqs.~(1) we derive the equation of motion for the  four-operator product $\hat{a}^+\hat{a}^+\hat{a}\hat{a}$ and  for its mean value $G_2 = \left<\hat{a}^+\hat{a}^+\hat{a}\hat{a}\right>$ we obtain the two coupled equations
\begin{subequations}\lb{7t}\beqr
    \dot{G}_2 & = & 4\Omega_0G_v - 4\kappa G_2, \lb{23}\\
\dot{G}_v & = &
\Omega_0\left[f\left( G_2 \Delta + 2N_en\right) + 0.5(\left<\hat{v}^+\hat{v}^+\hat{a}\hat{a}\right>+c.c.) \right.\nonumber \\
& & +
\left. 2\left<\hat{v}^+\hat{a}^+\hat{a}\hat{v}\right>'\right] - \left(3\kappa + \frac{\gamma_{\perp}}{2}\right)G_v,\lb{24}
\eeqr\end{subequations}
where $G_v \equiv \tfrac{1}{2}(\left<\hat{v}^+\hat{a}^+\hat{a}\hat{a}\right>+c.c.)$
and the prime means that operator products of the same emitters are excluded from sums in $\hat{v}^+\hat{v}$.
The steady-state solutions of Eqs.~\rf{7t} satisfy
\begin{subequations}\lb{SM5_01}\beqr
    G_v & = & \frac{\kappa}{\Omega_0 } G_2, \lb{SM5_011}\\
0 & = &
\Omega_0^2\left[f\left(\Delta G_2 + 2N_en\right) + \tfrac{1}{2}(\left<\hat{v}^+\hat{v}^+\hat{a}\hat{a}\right>+c.c.) \right.\nonumber \\
& + &
\left. 2\left<\hat{v}^+\hat{a}^+\hat{a}\hat{v}\right>'\right] - \left(3\kappa + \frac{\gamma_{\perp}}{2}\right)\kappa G_2,\lb{SM5_012}
\eeqr\end{subequations}
We approximate the four-operator mean values in Eq.~\rf{SM5_01} by products of two-operator mean values expressed through the stationary solution of Eqs.(2) of the main text. First, we replace  $\left<\hat{v}^+\hat{v}^+\hat{a}\hat{a}\right> \approx\left<\hat{v}^+\hat{a}\right>^2 $. Second, we note  that $\left<\hat{v}^+\hat{v}\right>' = \left<\hat{v}\hat{v}^+\right>'$ and that the mean value of the four-operator product  $\left<\hat{v}^+\hat{a}^+\hat{a}\hat{v}\right>'$ can be broken into the product of non-zero two-operator mean values in two different ways: as $\left<\hat{v}^+\hat{v}\right>'\left<\hat{a}^+\hat{a}\right>=fDn$ or as $\left<\hat{a}^+\hat{v}\right>\left<\hat{v}^+\hat{a}\right>$. We do not favor either of these two, adopting instead the symmetric truncation
for $\left<\hat{v}^+\hat{a}^+\hat{a}\hat{v}\right>'$
by approximating $\left<\hat{v}^+\hat{a}^+\hat{a}\hat{v}\right>' \approx \tfrac{1}{2}(fDn + \left<\hat{a}^+\hat{v}\right>\left<\hat{v}^+\hat{a}\right>)$. This procedure
is analogous to the cluster expansion technique used in~[3]. Inserting these approximations into Eq.~\rf{SM5_012} we obtain
\beqr
0 & = &
\Omega_0^2[2f N_en+ \tfrac{1}{2}\left(\left<\hat{v}^+\hat{a}\right>^2+\left<\hat{a}^+\hat{v}\right>^2 +2\left<\hat{a}^+\hat{v}\right>\left<\hat{v}^+\hat{a}\right>\right)
   \nonumber\\ & &
 + fDn ]  - \left(3\kappa + \frac{\gamma_{\perp}}{2}-\frac{\Omega_0^2f}{\kappa}\Delta \right)\kappa G_2 \nonumber\\[10pt]
 & = & \Omega_0^2\left[2f N_en+ \tfrac{1}{2}\Sigma^2+ fDn\right]  \nonumber\\ & & \hspace{2cm}
- \left[3\kappa + \frac{\gamma_{\perp}}{2}\left(1-\Delta/\Delta_{\rm th}\right) \right]\kappa G_2.\lb{SM5_21}
\eeqr
Here it has been used that $\Sigma=\left<v^+a\right>+\left<a^+v\right>$ and $\Delta_{\rm th}=\kappa\gamma_\perp/2\Omega_0^2f$. Thus,
\beq
		G_2 = \frac{2\Omega_0^2f(N_e+D)n-\Omega_0^2fDn+\tfrac{1}{2}(\Omega_0\Sigma)^2}{\kappa^2\left[3 + (\gamma_{\perp}/2\kappa)\left(1-\Delta/\Delta_{\rm th}\right)\right]}.\lb{SM5_03}
\eeq
From the stationary solution of Eqs.~(2) we obtain $\Omega_0\Sigma = 2\kappa n$, $D = (2\kappa/\gamma_{\perp}) n \Delta$ and
\beqr
2\Omega_0^2f(N_e+D) & = & \left(\kappa+\gamma_\perp/2\right)\Omega_0\Sigma-2\Omega_0^2f n \Delta  \nonumber\\[10pt]
	& = & \left(\frac{2\kappa}{\gamma_\perp} + 1-\frac{\Delta}{\Delta_{\rm th}}\right)\kappa\gamma_\perp n.	 \lb{SM5_04}	
\eeqr
Inserting these results into Eq.~\rf{SM5_03}, we find after a few lines of algebra
\beqr
G_2 & = & \frac{4+\tfrac{\gamma_\perp}{\kappa}\left(1-\Delta/\Delta_{\rm th}\right) - \Delta/\Delta_{\rm th}}{3+\tfrac{\gamma_\perp}{2\kappa}\left(1-\Delta/\Delta_{\rm th}\right)}\; n^2 \nonumber\\[10pt]
 & = &\left[1+\frac{\left(1+\tfrac{\gamma_\perp}{2\kappa}\right)\left(1-\Delta/\Delta_{\rm th}\right)}{3+\tfrac{\gamma_\perp}{2\kappa}\left(1-\Delta/\Delta_{\rm th}\right)}\right] n^2. \lb{SM5_061}
\eeqr
From the steady-state solution of Eqs.~(2) we can further deduce that
\beq
1 - \frac{\Delta}{\Delta_{\rm th}} = \frac{1+N_0/\Delta_{\rm th}}{2(1+2\kappa/\gamma_{\perp})n + 1},	\lb{SM5_07}	
\eeq
so that
\beq
G_2 = \left[1+\frac{\left(1+\tfrac{\gamma_\perp}{2\kappa}\right)\left(1+N_0/\Delta_{\rm th}\right)}{3+6n(1+2\kappa/\gamma_{\perp})+\tfrac{\gamma_\perp}{2\kappa}\left(1+N_0/\Delta_{\rm th}\right)}\right] n^2. \lb{SM5_08}
\eeq
This then leads to Eq.~(6) for the photon auto-correlation function $g_2 = G_2/n^2$.
%
%
\subsection{SM5: GLRE for alternative  laser level scheme  (with zero low-level population)}
%
%
In the main text we derived GLRE based on the assumption that the active medium consists of two-level emitters. Here we will assume another
level scheme, to show the generality of our approach and in order to illustrate that collective effects can affect lasers with different level schemes differently.

In particular, we
carry out  the  analogous analysis for a simplified but typical semiconductor laser scheme: we assume that the pump is from the manifold outside of the lasing transition, that pump blocking is negligible, and that the population $N_g$ of the lower level  of the lasing transition is approximately zero due to fast relaxation. Then instead of Eqs.~(2) we arrive at the following closed set of four coupled equations
\begin{subequations}\lb{SM6_t1}\hspace{-1cm}\beqr
    \dot{{n}} & = & \Omega_0{\Sigma} - 2\kappa {n},   \lb{SM6_1}\\
    \dot{{\Sigma}} & = & -(\gamma_{\perp}/2 + \kappa){\Sigma} + \nonumber\\
    & & \hspace{1.5cm} 2\Omega_0 f[{N}_e({n} + 1) + {D}],    \lb{SM6_2}\\
    \dot{{D}}  & = & -\gamma_{\perp} {D} + \Omega_0 {N}_e{\Sigma}, \lb{SM6_3}\\
    \dot{N}_e & = & - \Omega_0{\Sigma} + \gamma_{\parallel}(PN_0-N_e), \lb{SM6_4}
\eeqr\end{subequations}
from which the stationary photon output can be determined (as used below). Likewise, for the photon statistics, instead of Eqs.~\rf{7t} we obtain
\begin{subequations}\lb{SM6_t2}\beqr
    \dot{G}_2 & = & 4\Omega_0G_v - 4\kappa G_2, \lb{SM6_21}\\
\dot{G}_v & = &
\Omega_0\left[fN_e\left(G_2 + 2n\right) + \tfrac{1}{2}(\left<\hat{v}^+\hat{v}^+\hat{a}\hat{a}\right>+c.c.) \right.\nonumber \\
&  &
+ \left. 2\left<\hat{v}^+\hat{a}^+\hat{a}\hat{v}\right>'\right] - \left(3\kappa + \frac{\gamma_{\perp}}{2}\right)G_v.\lb{SM6_22}
\eeqr\end{subequations}
Let us now calculate the stationary $g_2$ based on this set of equations. Replacing as in SM4 the $\left<\hat{v}^+\hat{v}^+\hat{a}\hat{a}\right>$ by $\left<\hat{v}^+\hat{a}\right>^2$
and  $\left<\hat{v}^+\hat{a}^+\hat{a}\hat{v}\right>'$ by $\tfrac{1}{2}\left(fDn + \left<\hat{v}^+\hat{a}\right>\left<\hat{a}^+\hat{v}\right>\right)$,
we now obtain the identity
\beq
G_2 = \frac{2\Omega_0^2f(N_e+D)n-\Omega_0^2f Dn + (\Omega_0 \Sigma)^2/2}{\kappa^2 \left[3+ \tfrac{\gamma_{\perp}}{2\kappa}\left(1-N_e/N_{\rm th}\right)\right]},  \lb{SM6_3a}
\eeq
with $\kappa\gamma_{\perp}/(2\Omega_0^2f) \equiv N_{\rm th}$. From the stationary Eqs.~\rf{SM6_t1} we find $\Omega_0\Sigma = 2\kappa n$, $D = (2\kappa/\gamma_{\perp})N_e n$ and finally
\beq
2\Omega_0^2f(N_e+D) = \left[\frac{2\kappa}{\gamma_{\perp}}+1-\frac{N_e}{N_{\rm th}}\right]\kappa\gamma_{\perp} n. \lb{SM6_4a}
\eeq
Similar to Eq.~\rf{SM5_061} we can write
\beqr
	G_2 & = & \frac{\left(\frac{2\kappa}{\gamma_{\perp}}+1-\frac{N_e}{N_{\rm th}}\right)\kappa\gamma_{\perp} n^2-\kappa^2 \frac{N_e}{N_{\rm th}} n^2 + 2\kappa^2 n^2}{\kappa^2 \left[3+ \tfrac{\gamma_{\perp}}{2\kappa}\left(1-N_e/N_{\rm th}\right)\right]}\nonumber \\[10pt]
    & = & \frac{4+\frac{\gamma_{\perp}}{\kappa}\left(1-N_e/N_{\rm th}\right)-N_e/N_{\rm th}}{3+ \tfrac{\gamma_{\perp}}{2\kappa}\left(1-N_e/N_{\rm th}\right)} n^2 \nonumber \\[10pt]
    & = & \left[1+\frac{\left(1+\frac{\gamma_{\perp}}{2\kappa}\right)\left(1-N_e/N_{\rm th}\right)}{3+ \tfrac{\gamma_{\perp}}{2\kappa}\left(1-N_e/N_{\rm th}\right)}\right]n^2. \lb{SM6_3aa}
\eeqr
In fact, Eq.~\rf{SM6_3aa} can be obtained from the corresponding expression Eq.~\rf{SM5_061} in our two-level lasing model by replacing $\Delta$ with $N_e$ and renaming $\Delta_{\rm th}$ as $N_{\rm th}$. Inserting 	
\beq
	1 - \frac{N_e}{N_{\rm th}} = \frac{1}{1+ (1+2\kappa/\gamma_{\perp})n},	\lb{SM6_4b}	
\eeq
obtained from the steady state solutions to Eqs.~\rf{SM6_t1}, we find for $g_2 \equiv G_2/n^2$	
\beq
g_2(n) = 1+\frac{\gamma_{\perp}/2\kappa +1}{3+3n(1+2\kappa/\gamma_{\perp}) + \gamma_{\perp}/2\kappa}, \lb{SM6_50}
\eeq
where $n$ is the stationary solution of Eqs.~\rf{SM6_t1}:
\begin{subequations}
\begin{eqnarray}
		n & = & \frac{1}{2g}\left[\theta_0 +\left(\theta_0^2 + \frac{4g}{1+2\kappa/\gamma_{\perp}}\frac{N_0P}{N_{\rm th}}\right)^{1/2}\right] \lb{SM6_40} \\
\theta_0 & = & \frac{N_0P}{N_{\rm th}} - 1 - \frac{g}{1+2\kappa/\gamma_{\perp}}.
\end{eqnarray}
\end{subequations}
Now there are similarities and differences between the autocorrelation functions $g_{2}(n)$ of Eq.~(6) for the two-level laser and of Eq.~(\ref{SM6_50}) for the semiconductor laser model with infinitely fast ground-state depopulation. A similarity is that both decrease monotonically as a function of
$n$. A difference is that $g_{2}(n=0)$ of Eq.~(6) has $2+2\kappa/\gamma_{\perp}$ as an upper bound and  allows superthermal photon statistics as we presented in the main text, whereas $g_{2}(n=0)$ of Eq.~(\ref{SM6_50}) can be rewritten as $2 - 2/[3+ \gamma_{\perp}/(2\kappa)]$, clearly showing that superthermal statistics does not occur for this model. Thus the comparison of these two models indicates that for the build-up of superradiant correlations  we need at least a non-vanishing fraction of the emitters that make up the active medium to be in the lower lasing level.

Another difference between the two models is the dependence of $g_{2}$ on the number of emitters $N_{0}$. The curves in Fig.~3 of the main text show that  $g_2>2$ grows with $N_0$ at small $n$ for the two-level laser. Now turning to the semiconductor laser model, its corresponding $g_2$ of Eq.~(\ref{SM6_50}) depends on $N_0$ only implicitly through $n$, so that $d g_2/d N_0 = (\partial g_2/\partial n)(\partial n/\partial N_0)$. Because  $g_2$ decreases with $n$ (see Eq.~\rf{SM6_50}) and since $n$ increases with $N_0$, we now find instead that $g_{2}$ decreases with the number of emitters. The latter dependence is also seen in Fig.~5 of Ref.~[3]. So we find that the relative quantity $g_2$ can indeed vary with $N_0$ in different ways, depending on the laser scheme used.

Thus this SM5 illustrates that our approach to identify GLRE carries over to different laser level schemes, and that the importance of collective effects in the continuous-wave output of lasers will depend on the specific appropriate laser level schemes.